\input{aipcheck}
\documentclass[
    ,final            
]{aipproc}

\layoutstyle{6x9}

\begin{document}
\title{Spin dependent fragmentation functions at Belle}
\author{A. Ogawa}{
address={RIKEN Brookhaven Research Center \\
Upton, NY 11973-5000, USA\\
}
}
\author{M. Grosse-Perdekamp}{
address={University of Illinois at Urbana-Champaign \\
1100 W Green Street, \\ 
Urbana, IL 61801, USA\\ 
},
altaddress={RIKEN Brookhaven Research Center \\
Upton, NY 11973-5000, USA\\
}
}
\author{R. Seidl}{
address={University of Illinois at Urbana-Champaign \\
1100 W Green Street, \\ 
Urbana, IL 61801, USA\\ 
},
altaddress={RIKEN Brookhaven Research Center \\
Upton, NY 11973-5000, USA\\
}
}
\author{K. Hasuko}{
address={RIKEN \\
Wako, Saitama,351-0198, Japan\\
E-mail: mgp@uiuc.edu, akio@bnl.gov, rseidl@uiuc.edu\\
 for the Belle Collaboration}
}


\begin{abstract}
The measurement of the so far unknown chiral-odd quark transverse spin distribution in either semi-inclusive DIS (SIDIS) or inclusive measurements in pp collisions at RHIC has an additional chiral-odd fragmentation function appearing in the cross section. This chiral-odd fragmentation functions (FF) can for example be the so-called Collins FF \cite{collins} or the Interference FF. HERMES \cite{hermes} has given a first hint that these FFs are nonzero, however in order to measure the transversity one needs these FFs to be precisely known. 
We have used 29.0 fb$^{-1}$ of data collected by the Belle experiment at the KEKB $e^+e^-$ collider to measure azimuthal asymmetries for different charge combinations of pion pairs and thus access the Collins FF.\end{abstract}

\maketitle

\section*{Introduction}
At leading twist 3 quark distribution functions (DF) in the nucleon are present; the experimentally well measured unpolarized quark DF, the experimentally less  known quark helicity DF and the so far undetermined transversity DF. The latter cannot be measured in inclusive DIS due to its chiral-odd nature, since all possible interactions are chiral-even for nearly massless quarks. Therefore one needs an additional chiral-odd function in the cross section to access transversity. This can be achieved either by an anti quark transversity DF in double transversely polarized Drell-Yan processes or, alternatively, one can have a chiral-odd fragmentation function in semi-inclusive deep inelastic scattering (SIDIS) or hadroproduction.
\section{The Belle experiment}
The Belle \cite{belledetector} experiment at the asymmetric $e^+e^-$ collider KEKB at Tsukuba, Japan, is mainly dedicated to the study of CP violation in B meson decays. Its center of mass energy is tuned to the $\Upsilon(4S)$ resonance at $\sqrt{s} = 10.58$ GeV. Part of the data was also recorded 60 MeV below the resonance. These off-resonance events are studied in order to measure spin dependent fragmentation functions (FF). For the present analysis an integrated luminosity of 29.0 fb$^{-1}$ has been analyzed. The aerogel \v{C}erenkov counter, time-of-flight detector and the central drift chamber enable a good particle identification and tracking, which is crucial for these measurements. Using the information from the silicon vertex detector, one selects tracks originating from the interaction region and thus reducing the contribution of hadrons from heavy meson decays. To reduce the amount of hard gluon radiative events a cut on the kinematic variable thrust of $T> 0.8$ is applied. This enhances the typical 2-jet topology and the thrust axis is used as approximation of the original quark direction. We also require that the fractional energy $z \stackrel{CMS}{=} 2E_h/Q > 0.2$ in order to significantly reduce the contribution of pions arising from the vector meson decays or of those assigned to a wrong hemisphere.
\section*{Collins FF}
The Collins effect occurs in the fragmentation of a transversely polarized quark with polarization $S_q$ and 3-momentum k into an unpolarized hadron of transverse momentum $P_{h\perp}$ with respect to the original quark direction. According to the Trento convention \cite{trento} the number density for finding an unpolarized hadron h produced from a transversely polarized quark q is defined as:
\begin{equation}
D_{h\/q^\uparrow}(z,P_{h\perp})  = D_1^q(z,P_{h\perp}^2) + H_1^{\perp q}(z,P_{h\perp}^2)\frac{(\hat{\mathbf{k}} \times \mathbf{P}_{h\perp})\cdot \mathbf{S}_q}{zM_h} ,
\label{eq:cdef}
\end{equation}
where the first term describes the unpolarized FF $D^q_1(z,P_{h\perp})$, with $z\stackrel{CMS}{=} \frac{2E_h}{Q}$ being the fractional energy the hadron carries relative to half of the center of mass system (CMS) energy Q. The second term, containing the Collins function $H_1^{q\perp} (z,P^2_{h\perp})$, depends on the spin of the quark and thus leads to an asymmetry as it changes sign under flipping the quark spin. The vector product can accordingly be described by a $\sin(\phi)$ modulation, where $\phi$ is the azimuthal angle spanned by the transverse momentum and the plane defined by the quark spin and its momentum. In $e^+e^-$ hadron production the Collins effect can be observed by a combined measurement of a quark and an anti quark fragmentation. Combining two hadrons from different hemispheres in jetlike events, with azimuthal angles $\phi_1$ and $\phi_2$ as defined in Fig. 1, would result in a $\cos(\phi_1 +\phi_2)$ modulation. In the CMS these azimuthal angles are defined between the transverse component of the hadron momenta with regard to the thrust axis $\hat{n}$ and the plane spanned by the lepton momenta and $\hat{ n}$. The comparison of the thrust axis calculations using reconstructed and generated tracks in the MC sample shows an average angular separation between the two of 75 mrad with a a root mean square of 74 mrad. Due to that small biases in one of the reconstruction methods used could arise and were studied as discussed later. Following reference \cite{daniel} one either computes the azimuthal angles of each pion relative to the thrust axis which results in a $\cos(\phi_1 +\phi_2)$ modulation or one calculates the azimuthal angle relative to the axis defined by the $2^{nd}$ pion which results in a $\cos(2\phi_0)$ modulation. While the first method directly accesses moments of the Collins functions, the second method contains a convolution integral of the Collins FF over possible transverse momenta of the hadrons.
\begin{figure}

\includegraphics[width=4.8cm]{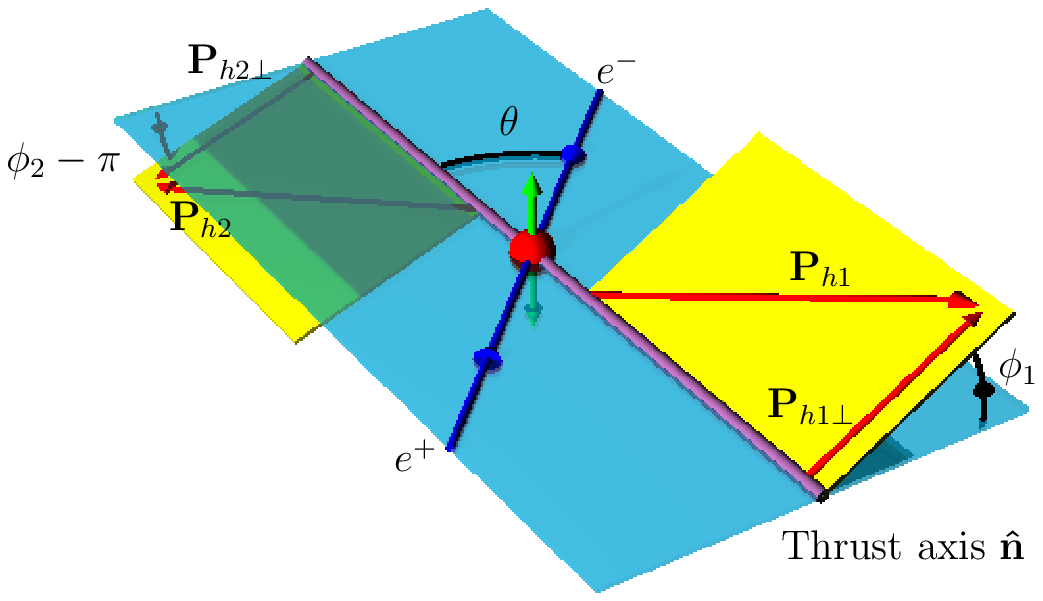}
\includegraphics[width=4cm]{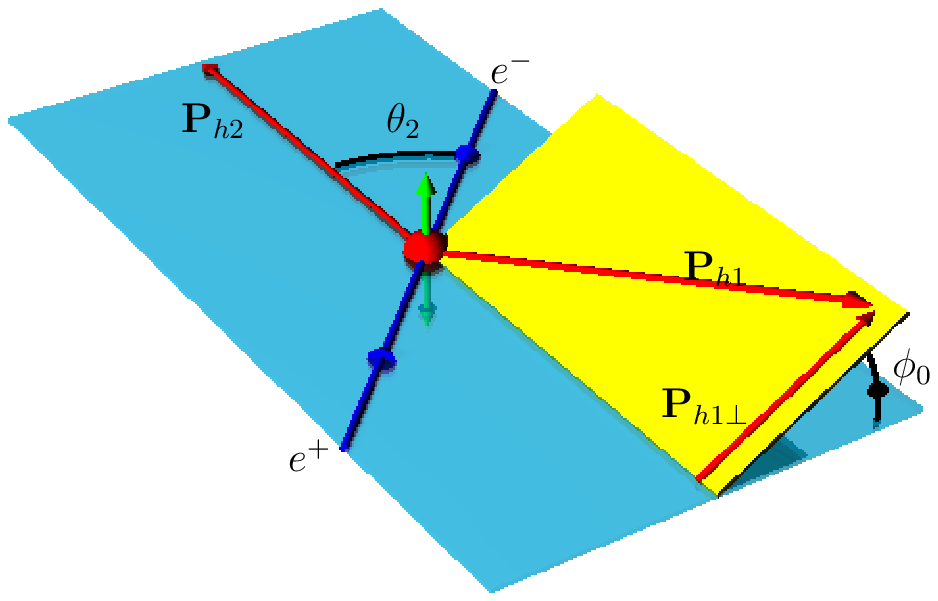}
\caption{Description of the azimuthal angles $\phi_0$, $\phi_1$ and $\phi_2$ relative to the scattering plane defined by the lepton axis and either the thrust axis  $\hat{n}$ or the momentum of the $2^{nd}$ hadron $P_{h2}$. \label{angle}}

\end{figure}
\subsection{Measured asymmetries}
We measure the azimuthal asymmetries $N(2\phi )/N_0$, where $N(2\phi )$ denotes the number of hadron pairs in bins of either $2\phi_0$ or $\phi_1 +\phi_2$ and $N_0$ is the average number of hadron pairs in the whole angle interval. The main backgrounds, producing similar azimuthal asymmetries as the Collins effect, are the radiation of soft gluons and possible acceptance effects. The gluonic contribution is proportional to the unpolarized FF and is independent of the charge of the hadrons. Consequently taking the ratio of the normalized distributions for unlike-sign over like-sign pairs the gluonic distributions cancel in the leading order:
\begin{eqnarray}
A_\alpha & := &\frac{\frac{N(2\phi )}{N_0} |_{unlikesign}}
{\frac{N(2\phi )}{N_0} |_{likesign}}  \nonumber \\
& \approx &  1+
\frac{\sin^2\theta}{1+cos^2\theta} \left( F\left(\frac{H_1^{\perp, fav}}{D_1^{fav}}
,\frac{H_1^{\perp,dis}}{D_1^{dis}}\right)+\mathcal{O} f(Q_T ,\alpha_S)^2\right)cos(2\phi ) \quad,
\end{eqnarray}
where $\theta$ is either the angle between the colliding leptons and the produced hadron or the thrust axis for methods $\alpha = 0, 12$, respectively. Favored and disfavored FF (fav,dis) describe the fragmentation of a light quark into a pion of same or opposite charge sign. Obviously also acceptance effects cancel in the double ratios. The latter are fit by the sum of a constant term and a $\cos(2\phi_0)$ or $\cos(\phi_1 +\phi_2)$ modulation. The double ratios of unlike sign over like sign pairs showed the existence of the Collins effect and gave a hint about the overall magnitude \cite{prl}. As suggested in \cite{peter}, measuring in addition double ratios containing any combination of charged pion pairs reveals additional information on the ratio of the favored and disfavored Collins functions. Preliminary results for the double ratios of unlike-sign (UL) over all charged (C) pion pairs can be seen in Fig.2 together with the final results of the unlike sign (UL) over like sign (L) pion pairs. The combined zbins are obtained by adding the symmetric bins of the $4\times 4$ $z \in$ [0.2,0.3,0.5,0.7,1.0] bins. The data has been corrected for the contribution of charmed hadron decays. A nonzero asymmetry is visible for both double ratios, while the UL/C are about 40\% of the UL/L results (the average values are: $A^{UL/C}_0 = (1.27 \pm 0.49 \pm 0.35)\%$ and $A^{UL/C}_{12} = (1.752 \pm 0.59 \pm 0.41)\%$ compared to $A^{UL/L}_0 = (3.06 \pm 0.57 \pm 0.55)\%$ and $A^{UL/L}_{12} = (4.26 \pm 0.68 \pm 0.68)\%$). The data shows a rising behavior with rising fractional energy z for both results. Several systematic cross-checks of the analysis method were performed and the differences in the results are quoted as systematic uncertainties: Instead of double ratios we used the subtraction method for the unlike from the like sign or charged pion asymmetries; the constant fit to the double ratios obtained in MC (without a Collins contribution) together with its statistical error and a similar fit to double ratios of positively charged over negatively charged pion pair data were assigned as systematic error. The differences to the results when fitting the double ratios also with higher order azimuthal modulations were added to the systematic errors. All contributions to the systematic errors were added in quadrature. Studies using introduced asymmetries in the MC data revealed that the $\cos(\phi_1+\phi_2)$ method undersetimates the generated asymmetries due to the discrepancies between the thrust axis calculated for generated and reconstructed tracks. These results have therefore been rescaled by a factor 1.21.
The presented measurement represents the first evidence of the Collins effect and will help to disentangle the favored to disfavored Collins function ratio.
\begin{figure}[ht]
\includegraphics[width=8cm]{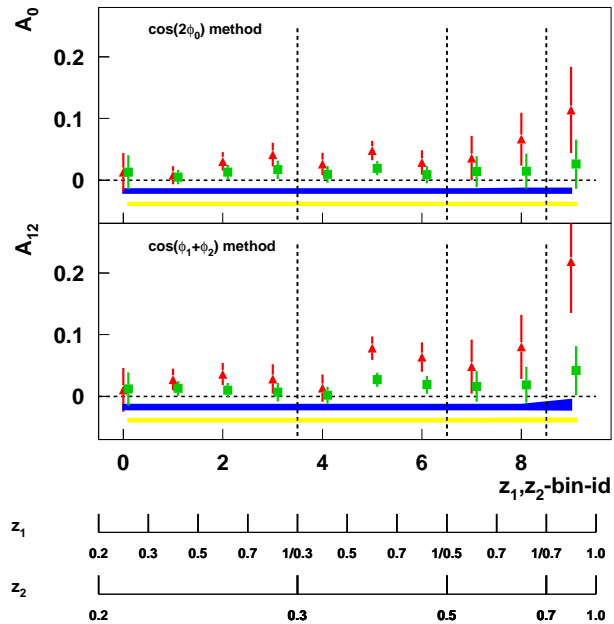}
\caption{Results for the $\cos(2\phi_0)$ and the $\cos(\phi_1+\phi_2)$ method for the UL/C (squares, preliminary) and UL/L (triangles, final) double ratios. The upper error band correspond to systematic errors of the UL/L double ratios, the lower one to those of the UL/C ratios.\label{fig:pipi}}
\end{figure}

%


\begin{thebibliography}{0}
\bibitem{collins}
J.~C.~Collins:
Nucl.~Phys~{\bf B396}(1993):161.
\bibitem{hermes}
A.~Airapetian et~al.(Hermes)
Phys.~Rev.~Lett.{\bf 94}(2005)012002.
\bibitem{belledetector}
A.~Abashian et~al.(Belle):
Nucl. Instrum. Meth.{\bf A479}(2002)117.
\bibitem{trento}
A.~Bacchetta, U.~D'Alesio, M.~Diehl, A.~Miller:
Phys.~Rev.~{\bf D70}(2004):117504.
\bibitem{daniel}
D.~Boer, R.~Jakob, P.~J.~Mulders:
Phys.~Let.~{\bf B 424}(1998):143.
\bibitem{prl}
R.~Seidl, et~al.(Belle):
Phys.~Rev.~Lett.~{\bf 96}(2006):232002.
\bibitem{peter}
A.~Efremov, K.~Goeke and P.~Schweitzer: [arXiv:hep-ph/0603054].



\end{thebibliography}
\end{document}